\documentclass{article}
\usepackage{iclr2027_conference,times}
\usepackage{hyperref}
\usepackage{url}

\protected\def\mysys{\textsc{OPFL}}

\usepackage[HTML]{xcolor}
\definecolor{softGreen}{HTML}{D0E0D0}

\newcommand{\secref}[1]{Sec.~\ref{#1}}

\usepackage{algorithm}
\usepackage{algpseudocode}
\usepackage{booktabs}
\usepackage{graphicx}
\usepackage{subcaption}
\usepackage{capt-of}

\usepackage{amsmath,amsfonts,bm}

\def\secref#1{section~\ref{#1}}

\def\eqref#1{equation~\ref{#1}}

\def\1{\bm{1}}

\DeclareMathAlphabet{\mathsfit}{\encodingdefault}{\sfdefault}{m}{sl}
\SetMathAlphabet{\mathsfit}{bold}{\encodingdefault}{\sfdefault}{bx}{n}

\title{OPFL: Optimistic Verification of Federated Learning via Empirical Boundary}
\author{Hongxu Su \\
\texttt{hsu238@connect.hkust-gz.edu.cn}
\And
Jianzhu Yao \\
\texttt{jy0246@princeton.edu}
\AND
Xuechao Wang \\
\texttt{xuechaowang@hkust-gz.edu.cn}
\And
Pramod Viswanath \\
\texttt{pramodv@princeton.edu}
}

\iclrfinalcopy
\hypersetup{
  pdftitle={OPFL: Optimistic Verification of Federated Learning via Empirical Boundary},
  pdfauthor={Hongxu Su, Jianzhu Yao, Xuechao Wang, Pramod Viswanath}
}

\begin{document}
\maketitle
\fancyhead{}
\renewcommand{\headrulewidth}{0pt}

\begin{abstract}
Federated learning enables multiple clients to collaboratively train models without sharing their private data. However, the lack of visibility into local training makes it difficult to verify whether clients follow the prescribed training procedure or submit malicious updates, such as model poisoning. 
A natural approach is to replay client training for verification. However, privacy-preserving replay produces numerical results that cannot be directly matched with local client execution because the two run in different environments. We present \mysys{}, an optimistic verification framework for privacy-preserving federated learning. To protect data privacy, \mysys{} performs replay inside secure multi-party computation (MPC). Although gradients computed on MPC and local GPUs are not bitwise identical, we observe that their absolute differences are stable and bounded. \mysys{} therefore calibrates an empirical boundary offline and uses it to distinguish benign numerical deviations from malicious manipulation. To reduce the cost of expensive MPC replay, \mysys{} adopts optimistic verification by post auditing only sampled training steps. 
Experiments on LeNet, BERT, and Qwen show that the boundary generalizes across datasets, input lengths, and GPUs, while achieving $0$\% ASR against model poisoning and PGD-based attacks. On a LeNet workload, at $p=0.01$, \mysys{} is approximately $98.6\times$ faster than full MPC-based FL and $625.5\times$ faster than ZK-based approach.
\end{abstract}

\section{Introduction}

Federated learning (FL) has become an important paradigm for collaborative machine learning in privacy-sensitive settings, including on-device learning and multi-institutional healthcare~\citep{mcmahan2017communication,molaei2024federated}. Instead of collecting raw data at a central server, FL allows multiple clients to train a global model using their local datasets. In each round, clients download the current global model, perform local training, and return only model updates to the server for aggregation~\citep{mcmahan2017communication}. The raw training data therefore remains on client devices. However, this privacy benefit also makes local training difficult to observe. A malicious client may deviate from the prescribed training procedure or directly construct a poisoned update while still submitting it as the result of legitimate local training~\citep{bagdasaryan2020backdoor,zhuang2024backdoor,pang2025poisafl}. This creates an execution integrity problem: how can a server verify that a private client update was actually produced by the prescribed local training procedure?

Existing approaches address this problem from several directions. 
Secure aggregation with update validation can constrain which client updates are accepted without revealing them~\citep{lycklama2023rofl,chowdhury2022eiffel,bell2023acorn,zhu2024risefl}. 
However, these methods primarily verify properties or constraints of the submitted update rather than directly establishing how the update was produced.
Cryptographic approaches can verify training computations more directly. 
For example, recent systems use zero-knowledge proofs to enforce private computation integrity in federated or split learning settings~\citep{zhu2024risefl,zheng2026zksl}. 
Nevertheless, encoding deep neural-network training into cryptographic proofs remains expensive and difficult to scale~\citep{zhu2024risefl,zheng2026zksl}.
Replay provides another natural alternative. 
Prior approaches to verifiable training record intermediate training states and re-execute selected computations to check consistency with the claimed training trajectory~\citep{jia2021prooflearning,srivastava2024optimistic}. 
Yet replay is difficult to apply directly in federated learning. A verifier would need to reproduce the client's execution environment as closely as possible and access the raw training data used in local computation, which conflicts with the privacy goal of federated learning. A natural alternative is to perform replay inside a cryptographically protected environment such as MPC. However, such environments differ from native GPU execution and can introduce unavoidable numerical discrepancies~\citep{goldberg1991floating,demmel2015parallel,collange2015numerical,srivastava2024optimistic}. Their outputs cannot generally be compared through exact equality, making direct verification ineffective.

To address these challenges, we introduce \mysys{}, a framework for privately verifying FL through an empirical gradient-discrepancy boundary. \mysys{} uses MPC to enable \textbf{privacy-preserving replay}, but does not require the replayed gradient to be bitwise identical to the gradient produced by native GPU training. Our key observation is that the gradient differences between GPU execution and MPC replay exhibit highly structured behavior and remain empirically bounded. This structured behavior gives rise to an \textbf{empirical boundary} that separates benign numerical discrepancies from malicious deviations.
\mysys{} calibrates this boundary by executing the same training configuration in both native GPU and MPC environments and measuring the absolute gradient discrepancy \(\lvert G_{\mathrm{GPU}} - G_{\mathrm{MPC}} \rvert\) in limited training steps. The resulting discrepancy distribution is summarized using a subset of percentile statistics, which preserves distributional information while avoiding the cost of checking every gradient coordinate. This calibration is performed offline once before FL starts and is reused during subsequent verification. 
Replaying every training step inside MPC is computationally expensive~\citep{mohassel2018aby3,ma2023spu}. To reduce this cost, \mysys{} adopts an \textbf{optimistic auditing mechanism} and invokes MPC replay only on sampled training steps. This design follows a well-established paradigm in optimistic verification systems~\citep{kalodner2018arbitrum}. Claims are accepted optimistically, while detected misbehavior can trigger challenges and economic penalties. Clients commit their training evidence before audit sampling. A verifier committee then replays only the selected steps inside MPC and checks the resulting gradient against the calibrated boundary. Failed audits trigger deposit slashing, under the assumption that a majority of committee members are honest.

We evaluate \mysys{} across diverse training settings, including CNN and Transformer architectures, multiple optimizers, and both full model training and LoRA finetuning~\citep{hu2022lora}.
First, we evaluate whether \mysys{} can distinguish honest numerical discrepancies from malicious deviations. We consider four representative attacks adapted from prior work, including gradient reuse, gradient sign reversal, label flipping, and gradient scaling~\citep{fraboni2021freerider,damaskinos2018byzantine,fang2020local,bagdasaryan2020backdoor,lycklama2023rofl}. \mysys{} achieves 0\% ASR while maintaining 0\% false rejection rate across these attacks. We further construct a white-box adaptive PGD attack. PGD is widely regarded as a strong first-order adversary~\citep{madry2018towards}, making it a suitable stress test for our verification framework. The ASR remains 0\% for \mysys{}. In contrast, the evaluated baselines can detect large deviations but are more easily bypassed under smaller adaptive perturbations. Detailed results are provided in \secref{sec:evaluation}. 
Second, we study the stability and generalization of the calibrated boundary. As shown in \secref{sec:boundary-generalization}, a stable boundary can be obtained from only a small number of calibration steps. Across LeNet, BERT, and Qwen3, boundaries calibrated using the first 100 training steps generalize across different training stages, datasets, sequence lengths, and GPU hardware.
Finally, we evaluate the system overhead of \mysys{}. In our five-client LeNet workload, a 1\% audit rate reduces the measured verification and aggregation time by approximately $98.6\times$ relative to full MPC-based verification and $625.5\times$ relative to ZKSL-LP-G~\citep{zheng2026zksl}.

In summary, our key finding is that the gradient discrepancy between MPC execution and native GPU execution is bounded, and can therefore serve as an effective signal for identifying malicious training deviations, as demonstrated by our experiments. We further build a verification framework that audits FL clients without exposing their private training data, and introduce an optimistic verification mechanism to reduce system overhead. \mysys{} focuses on execution integrity and does not address data poisoning, which can be handled by complementary robust aggregation or trust-based defenses~\citep{blanchard2017krum,cao2021fltrust}. The construction and incentivization of the verifier committee are also orthogonal to our design and can rely on established decentralized committee-selection and incentive mechanisms~\citep{gilad2017algorand,kiayias2017ouroboros}. Finally, \mysys{} assumes an existing privacy-preserving aggregation layer, such as secure aggregation~\citep{bonawitz2017secureaggregation}, rather than introducing a new aggregation protocol.
\section{Problem Formulation and Threat Model}
\label{sec:problem}

\paragraph{Motivation.}
FL assumes that participating clients faithfully execute the prescribed local training procedure before submitting their updates. In practice, a malicious or economically motivated client may deviate from this procedure. For example, it may skip computation, reuse a stale gradient, or directly fabricate an update to reduce local training cost. A client may also manipulate its computation to steer the global model toward a malicious objective~\citep{fraboni2021freerider,bagdasaryan2020backdoor}. Since the server observes only the submitted update, such deviations are difficult to distinguish from honestly generated updates without additional verification.

\paragraph{FL Setting.}
\label{sec:fl-setting}

We consider a FedSGD-style FL setting~\citep{mcmahan2017communication} with a central server $\mathcal{S}$ and a set of clients $\{\mathcal{C}_i\}_{i=1}^{n}$. Each client $\mathcal{C}_i$ holds a private local dataset $D_i$. At training round $t$, the server distributes the current global model $W_t$ to the selected clients. Let $z_{i,t}$ denote the private training input used by client $\mathcal{C}_i$ at round $t$, derived from $D_i$. In this setting, each selected client performs a single local training step per round and submits the resulting gradient for aggregation. We use this formulation for clarity; the same verification principle can be extended to multi-step local training by treating each local step as an auditable computation and binding the corresponding intermediate states.
Each selected client performs the prescribed local computation according to a public training configuration $\Pi$, which specifies the training algorithm, hyperparameters, model and optimizer settings, data-processing rules, randomness configuration, and other metadata required for replay. Client $\mathcal{C}_i$ computes

\begin{equation}
G_{i,t}^{\mathrm{GPU}}
=
\operatorname{Train}_{\mathrm{GPU}}
\left(
W_t,z_{i,t};\Pi
\right),
\end{equation}

and submits the resulting gradient for aggregation.

We augment this setting with a verifier committee $\mathcal{K}$ that audits randomly selected client-round pairs $(i,t)$. Before audit selection, each client commits to its private training inputs and claimed gradients. The global model $W_t$ and training configuration $\Pi$ are fixed independently of the client. We assume a public and binding commitment layer that prevents committed evidence from being modified.
Our goal is not to determine whether a client's private dataset is itself benign. Instead, we verify whether the committed gradient is consistent with executing the prescribed local computation on the committed private input.

\paragraph{Verification Goal.}
\label{sec:verification-goal}

A direct way to verify a client gradient is to replay the same computation and compare the result with the client's claim. However, revealing $z_{i,t}$ to the verifier would violate the privacy objective of FL. We therefore perform replay inside secure multi-party computation (MPC)~\citep{mohassel2018aby3,ma2023spu}. For an audited client-round pair $(i,t)$, the committee obtains

\begin{equation}
G_{i,t}^{\mathrm{MPC}}
=
\operatorname{Train}_{\mathrm{MPC}}
\left(
W_t,z_{i,t};\Pi
\right)
\end{equation}

without revealing $z_{i,t}$ to the committee members.

Exact replay would require
$G_{i,t}^{\mathrm{GPU}}=G_{i,t}^{\mathrm{MPC}}$.
In practice, this requirement is too strong. Native GPU execution and MPC replay use different numerical representations and computation environments, and can therefore produce different gradients even when both executions are honest~\citep{srivastava2024optimistic}.

We therefore verify the discrepancy between the two gradients rather than requiring exact equality. For gradient coordinate $j$, the absolute discrepancy is

\begin{equation}
\Delta_{i,t,j}^{\mathrm{abs}}
=
\left|
\left[G_{i,t}^{\mathrm{GPU}}\right]_j
-
\left[G_{i,t}^{\mathrm{MPC}}\right]_j
\right|.
\end{equation}

Let $\mathcal{B}$ denote the empirical gradient-discrepancy boundary calibrated from honest GPU--MPC executions. An audited client-round pair is accepted if its discrepancy profile satisfies the thresholds specified by $\mathcal{B}$, and rejected otherwise. Section~\ref{sec:design} defines the complete absolute and relative discrepancy profiles and describes how $\mathcal{B}$ is calibrated.

Accordingly, \mysys{} targets two properties. First, \emph{privacy}: verification should not reveal the client's raw training input to the verifier committee. Second, \emph{execution integrity}: an audited deviation that produces discrepancies beyond the calibrated numerical tolerance should be rejected.

\paragraph{Threat Model and Assumptions.}
\label{sec:threat-model}

We consider malicious clients that may arbitrarily deviate from the prescribed local computation by skipping computation, reusing previous gradients, modifying intermediate computation, or submitting arbitrary gradients. The adversary knows the verification procedure and boundary $\mathcal{B}$ and may adapt its strategy accordingly.

We assume:
\begin{itemize}
    \item \textbf{Post-commitment auditing.} Client evidence is committed before audit randomness is revealed, while $W_t$, $\Pi$, and $\mathcal{B}$ are fixed independently of the client.
    \item \textbf{Binding commitments.} Committed inputs and gradients cannot be replaced without detection.
    \item \textbf{MPC security.} Committee members follow the MPC protocol, and corrupted members remain within its privacy threshold, preserving input confidentiality and correct verification.
\end{itemize}

\paragraph{Scope.} \mysys{} focuses on verifying the integrity of local training execution. It does not prevent data poisoning when a client faithfully executes the prescribed computation on malicious or poisoned local data; such threats can be addressed by complementary robust aggregation or trust-based defenses~\citep{blanchard2017krum,cao2021fltrust}. We also do not address the construction of the verifier committee, which can rely on established decentralized committee-selection mechanisms~\citep{gilad2017algorand,kiayias2017ouroboros}. Finally, \mysys{} assumes an existing privacy-preserving aggregation mechanism, such as secure aggregation~\citep{bonawitz2017secureaggregation}, rather than introducing a new aggregation protocol. These components are orthogonal to our work.
\section{\mysys{} Design}
\label{sec:design}

\begin{figure}[htbp]
    \centering
    \includegraphics[width=\linewidth]{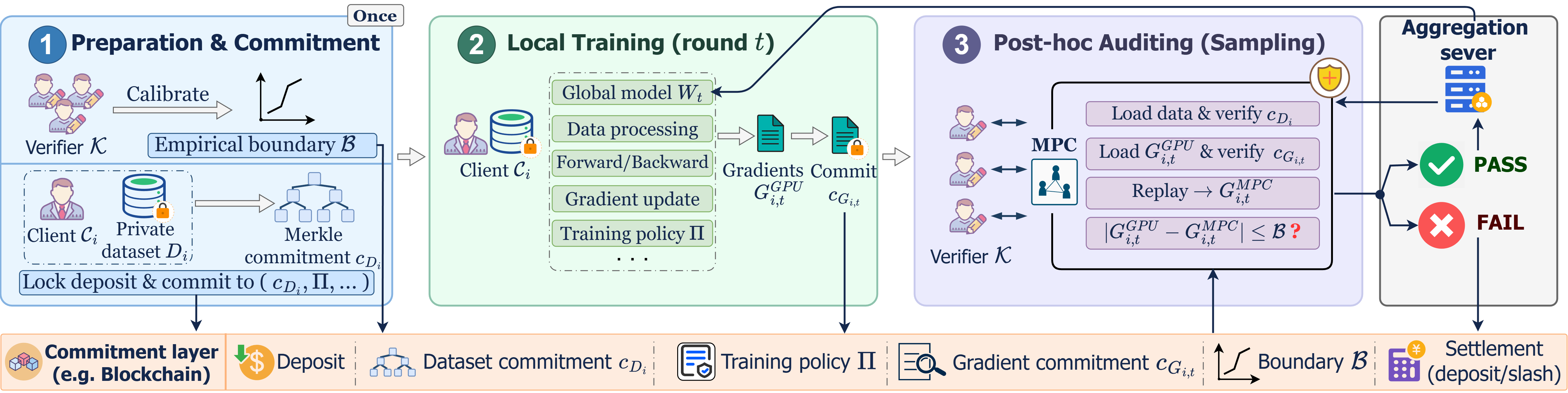}
    \caption{Overview of the \mysys{} protocol.}
    \label{fig:ovfed-overview}
\end{figure}

\mysys{} operates in three stages: preparation and commitment, local training, and pre-aggregation verification. The first stage fixes the verification rule and training configuration before execution. The second stage commits the client gradients produced during FL and sends them privately to the verifier committee. The third stage verifies every gradient against its commitment and performs MPC replay on randomly sampled client-round pairs before eligible contributions are released for aggregation. Figure~\ref{fig:ovfed-overview} provides an overview of the workflow.

\paragraph{Preparation and commitment.}

\mysys{} relies on a public and binding commitment layer to record deposits and protocol commitments. A natural instantiation is a blockchain, which provides an immutable record and supports programmable deposit-and-slashing rules. Each client locks a predefined deposit on the blockchain.

Before FL begins, the verifier performs offline calibration using a set of honest calibration steps. Let $s$ index a calibration step and $j$ a gradient coordinate. We define the absolute and relative discrepancies as
\begin{equation}
\Delta_{s,j}^{\mathrm{abs}}
=
\left|
[G_s^{\mathrm{GPU}}]_j
-
[G_s^{\mathrm{MPC}}]_j
\right|,
\qquad
\Delta_{s,j}^{\mathrm{rel}}
=
\frac{
\Delta_{s,j}^{\mathrm{abs}}
}{
\max\{
|[G_s^{\mathrm{GPU}}]_j|,
|[G_s^{\mathrm{MPC}}]_j|
\}
+\epsilon
}.
\end{equation}
Here, $\epsilon>0$ prevents division by zero and stabilizes the relative discrepancy near zero.

For each calibration step $s$, the verifier treats
$\{\Delta_{s,j}^{\mathrm{abs}}\}_{j=1}^{d}$ and
$\{\Delta_{s,j}^{\mathrm{rel}}\}_{j=1}^{d}$
as empirical discrepancy distributions over the $d$ gradient coordinates.
Let $Q_s^{\mathrm{abs}}(p)$ and $Q_s^{\mathrm{rel}}(p)$ denote their empirical $p$-quantiles. We evaluate these quantiles on
$\mathcal{P}=\{0.02,0.05,0.10,0.15,\ldots,0.90,0.95,0.98\}$.
The grid $\mathcal{P}$ can be adjusted to trade verification granularity for cost. A denser grid captures more distributional information and stricter but requires more quantile computation and boundary checks.

Percentile profiles characterize the overall discrepancy distribution but leave extreme tail coordinates unconstrained. We therefore additionally define
$\Delta_s^{\infty}
=
\|G_s^{\mathrm{GPU}}-G_s^{\mathrm{MPC}}\|_{\infty}$.
Let $\mathcal{S}_{\mathrm{cal}}$ denote the set of calibration steps. The raw bounds are
\begin{equation}
\widetilde{B}_{p}^{\mathrm{abs}}
=
\max_{s\in\mathcal{S}_{\mathrm{cal}}}
Q_s^{\mathrm{abs}}(p),
\quad
\widetilde{B}_{p}^{\mathrm{rel}}
=
\max_{s\in\mathcal{S}_{\mathrm{cal}}}
Q_s^{\mathrm{rel}}(p),
\quad
\widetilde{B}_{\infty}
=
\max_{s\in\mathcal{S}_{\mathrm{cal}}}
\Delta_s^{\infty}.
\end{equation}

To provide additional tolerance for deployment-time numerical variation, the raw bounds are enlarged by a predefined safety factors $\alpha_{\mathrm{abs}}, \alpha_{\mathrm{rel}}, \alpha_{\infty} \in \mathbb{R}_{>0}$: 
\begin{equation}
B_{p}^{\mathrm{abs}}
=
\alpha_{\mathrm{abs}}\widetilde{B}_{p}^{\mathrm{abs}},
\quad
B_{p}^{\mathrm{rel}}
=
\alpha_{\mathrm{rel}}\widetilde{B}_{p}^{\mathrm{rel}},
\quad
B_{\infty}
=
\alpha_{\infty}\widetilde{B}_{\infty}.
\end{equation}

The deployed verification boundary is $\mathcal{B}=\left(\{B_p^{\mathrm{abs}},B_p^{\mathrm{rel}}\}_{p\in\mathcal P},B_{\infty}\right)$.

Boundary calibration is performed only once for a given model and the resulting boundary is fixed before FL begins. The client then commits to the private training inputs used during training. Let
$\widetilde{D}_i=(z_{i,0},\ldots,z_{i,N_i-1})$
denote the canonicalized sequence of training inputs of client $\mathcal{C}_i$. For each input, the client samples private randomness $r_{i,t}$. The input commitment and dataset root are
\begin{equation}
c_{i,t}
=
H\!\left(
\mathsf{tag}_D
\Vert r_{i,t}
\Vert \operatorname{Encode}(z_{i,t})
\right),
\qquad
c_{D_i}
=
\mathsf{MerkleRoot}
(c_{i,0},\ldots,c_{i,N_i-1}),
\end{equation}
where $\operatorname{Encode}(\cdot)$ denotes a deterministic canonical encoding.

The dataset commitment $c_{D_i}$, calibrated boundary $\mathcal{B}$, and training policy $\Pi$ are fixed and publicly recorded on the commitment layer before training, while the underlying dataset remains private.

\paragraph{Local training and gradient commitment.}

At round $t$, client $\mathcal{C}_i$ uses the global model $W_t$, the committed training batch $z_{i,t}$, and public policy $\Pi$ to compute
$G_{i,t}^{\mathrm{GPU}}=\operatorname{Train}_{\mathrm{GPU}}(W_t,z_{i,t};\Pi)$.

The client samples private randomness $r_{G_{i,t}}$ and publishes the gradient commitment
$c_{G_{i,t}}
=
H(\mathsf{tag}_G
\Vert r_{G_{i,t}}
\Vert
\operatorname{Encode}(G_{i,t}^{\mathrm{GPU}}))$
before audit selection. Instead of sending the gradient directly to the aggregation server, the client secret-shares
$G_{i,t}^{\mathrm{GPU}}$ and $r_{G_{i,t}}$
to the verifier committee for pre-aggregation verification.

\paragraph{Pre-aggregation verification and auditing.}

Verification proceeds in two gates before aggregation. First, every submitted gradient is checked inside MPC against its previously published commitment. For client-round pair $(i,t)$, the committee verifies
\begin{equation}
b_{G,i,t}
=
\left[
H\!\left(
\mathsf{tag}_G
\Vert r_{G_{i,t}}
\Vert
\operatorname{Encode}(G_{i,t}^{\mathrm{GPU}})
\right)
=
c_{G_{i,t}}
\right].
\end{equation}
A contribution that fails this commitment check is rejected and never enters aggregation.

Second, after gradient commitments are fixed, the protocol randomly samples client-round pairs for replay auditing. An unaudited contribution that passes the commitment check is accepted optimistically. For an audited pair $(i,t)$, the client additionally secret-shares the committed training input $z_{i,t}$ and its opening $r_{i,t}$. The committee verifies that the input belongs to the committed dataset and then replays the prescribed computation inside MPC:
$G_{i,t}^{\mathrm{MPC}}=\operatorname{Train}_{\mathrm{MPC}}(W_t,z_{i,t};\Pi)$.

The committee computes the absolute and relative discrepancy profiles between
$G_{i,t}^{\mathrm{GPU}}$ and $G_{i,t}^{\mathrm{MPC}}$.
Let $P_{i,t}^{\mathrm{abs}}(p)$ and
$P_{i,t}^{\mathrm{rel}}(p)$ denote the corresponding empirical
$p$-quantiles. It also computes the tail discrepancy
$\Delta_{i,t}^{\infty}
=
\|G_{i,t}^{\mathrm{GPU}}
-
G_{i,t}^{\mathrm{MPC}}\|_{\infty}$.
The replay audit passes only if
\begin{equation}
P_{i,t}^{\mathrm{abs}}(p)
\leq B_p^{\mathrm{abs}},
\quad
P_{i,t}^{\mathrm{rel}}(p)
\leq B_p^{\mathrm{rel}},
\quad
\forall p\in\mathcal P;
\qquad
\Delta_{i,t}^{\infty}
\leq B_{\infty}.
\end{equation}

Only contributions that pass the commitment gate and, when sampled, the replay audit are authorized for aggregation. The committee forwards the same verified gradient shares to the aggregation layer, so the client does not provide a second aggregation input. Contributions that fail either gate are excluded from aggregation and trigger the predefined penalty. Only the final eligibility decision is revealed.

Let $a_{i,t}\in\{0,1\}$ denote the post-commitment audit indicator, where $a_{i,t}=1$ indicates that $(i,t)$ is selected for replay. Algorithm~\ref{alg:preagg-verification} summarizes the procedure.

\begin{algorithm}[htbp]
\caption{Pre-aggregation verification in \mysys{}}
\label{alg:preagg-verification}
\begin{algorithmic}[1]

\Require $W_t,\Pi,\mathcal{B},c_{D_i},c_{G_{i,t}},c_{i,t},a_{i,t}$
\Require Private $[G_{i,t}^{\mathrm{GPU}}],[r_{G_{i,t}}]$
\Require If $a_{i,t}=1$: private $[z_{i,t}],[r_{i,t}]$; public Merkle path $\pi_{i,t}$
\Ensure $y_{i,t}\in\{\mathsf{PASS},\mathsf{FAIL}\}$

\State $[b_G]\gets
[H(\mathsf{tag}_G\Vert[r_{G_{i,t}}]
\Vert\operatorname{Encode}([G_{i,t}^{\mathrm{GPU}}]))
=c_{G_{i,t}}]$

\If{$a_{i,t}=1$}

\State $[b_D]\gets
[H(\mathsf{tag}_D\Vert[r_{i,t}]
\Vert\operatorname{Encode}([z_{i,t}]))=c_{i,t}]
\land
\mathsf{MerkleVerify}(c_{i,t},\pi_{i,t},c_{D_i})$

\State $[G_{i,t}^{\mathrm{MPC}}]\gets
\operatorname{Train}_{\mathrm{MPC}}
(W_t,[z_{i,t}];\Pi)$

\State $[P_{i,t}^{\mathrm{abs}}],
[P_{i,t}^{\mathrm{rel}}],
[\Delta_{i,t}^{\infty}]
\gets
\mathsf{DiscrepancyProfile}
([G_{i,t}^{\mathrm{GPU}}],
[G_{i,t}^{\mathrm{MPC}}])$

\State $[b_{\mathcal B}]\gets
\displaystyle\bigwedge_{p\in\mathcal P}
\left(
[P_{i,t}^{\mathrm{abs}}(p)\leq B_p^{\mathrm{abs}}]
\land
[P_{i,t}^{\mathrm{rel}}(p)\leq B_p^{\mathrm{rel}}]
\right)
\land
[\Delta_{i,t}^{\infty}\leq B_{\infty}]$

\State $[y_{i,t}]
\gets
[b_G]\land[b_D]\land[b_{\mathcal B}]$

\Else

\State $[y_{i,t}]\gets[b_G]$

\EndIf

\State $y_{i,t}\gets\mathsf{Open}([y_{i,t}])$
\State \Return $y_{i,t}$

\end{algorithmic}
\end{algorithm}
\section{Experimental Evaluation}
\label{sec:evaluation}

We evaluate \mysys{} on three representative models with different scales and training settings, as summarized in Table~\ref{tab:models}. Our evaluation covers both CNN and Transformer architectures, multiple optimizers, and both full model training and LoRA finetuning. 
The attack experiments evaluate the verification rule of \mysys{} independently of audit sampling and assume that all tested instances are audited. System-level security further relies on optimistic auditing with deposit-and-slashing incentives.

\begin{table}[htb]
\centering

\setlength{\tabcolsep}{3.5pt} 
\caption{Models and training configurations used in our evaluation.}
\label{tab:models}
\begin{tabular}{lcccc}
\toprule
\textbf{Model} & \textbf{\#Parameters} & \textbf{Training} & \textbf{Optimizer} & \textbf{Datasets} \\
\midrule
LeNet & 431K & Full training & Momentum SGD & MNIST, Fashion-MNIST \\
BERT-Base & 110M & LoRA, last two layers & AdamW & SST-2, CoLA \\
Qwen3 & 0.6B & LoRA, last two layers & SGD & Alpaca, Dolly \\
\bottomrule
\end{tabular}
\end{table}

Our evaluation focuses on three questions. First, because the empirical boundary is the central verification criterion in \mysys{}, we study its stability and generalization. In particular, we examine whether a reliable boundary can be calibrated from only a small number of training steps and subsequently generalized across different training conditions, reducing the cost of the calibration stage. Second, we evaluate the effectiveness of \mysys{} against malicious training deviations. We consider gradient reuse, gradient sign reversal, label flipping, gradient scaling, and an adaptive PGD-based attack designed to evade the calibrated boundary. We further compare \mysys{} with RoFL-$L_2$, RoFL-$L_\infty$, EIFFeL, and RiseFL by evaluating whether each method can reject the same adaptive malicious updates under its respective verification rule. Finally, we evaluate the system overhead of \mysys{} and compare its cost with full MPC-based FL and ZK-based verification.

\paragraph{Experimental setup.}
We implement \mysys{} in PyTorch with Secure Processing Unit (SPU)~\citep{ma2023spu} as the MPC backend under a three-party semi-honest setting. Experiments use four NVIDIA RTX 4090 GPUs and two Intel Xeon Platinum 8336C CPUs at 2.30\,GHz, while cross-device generalization is evaluated on an additional RTX A4500 server.

\subsection{Effectiveness against model poisoning}
\label{sec:effectiveness}

\paragraph{General attacks.}

We first evaluate whether \mysys{} can reject malicious updates while still accepting honest computation. Reporting ASR alone is insufficient, since a verifier that rejects every update would trivially achieve 0\% ASR. We therefore report both ASR and false rejection rate (FRR). ASR is the fraction of attacked steps that both evade verification and achieve their intended malicious effect, while FRR is the fraction of unattacked honest steps that are incorrectly rejected:
\begin{equation}
\mathrm{FRR}
=
\frac{\#\text{ rejected honest steps}}
{\#\text{ honest verification steps}}.
\end{equation}

We consider four representative attack patterns adapted from prior work on free-riding, Byzantine gradient manipulation, data poisoning, and model-replacement attacks in federated and distributed learning~\citep{fraboni2021freerider,damaskinos2018byzantine,fang2020local,bagdasaryan2020backdoor,lycklama2023rofl}.
For each model, we evaluate 2,000 training steps and randomly select 20\% of them for attack. Let $G_{i,t}^{\mathrm{GPU}}$ denote the honest client gradient defined previously, and let $\widetilde{G}_{i,t}$ denote the gradient submitted under attack. We consider: (i) \emph{Gradient Reuse}, where the attacker computes a fresh honest gradient once every $k$ steps and reuses the most recent gradient in the intermediate steps, with replay interval $k\in\{2,5,10\}$; (ii) \emph{Reverse}, where $\widetilde{G}_{i,t}=-\gamma G_{i,t}^{\mathrm{GPU}}$ with $\gamma\in\{0.5,1,2\}$; (iii) \emph{Label Flip}, where the supervision target is replaced while keeping the model state and input unchanged, and the resulting gradient is submitted; and (iv) \emph{Amplify}, where $\widetilde{G}_{i,t}=\gamma G_{i,t}^{\mathrm{GPU}}$ with $\gamma\in\{5,10\}$. 
We evaluate these attacks on LeNet, BERT-Base, and Qwen3-0.6B. As summarized in Table~\ref{tab:attack-summary}(a), \mysys{} achieves 0\% ASR and 0\% FRR across different settings.

\paragraph{PGD-based adaptive attack.}

We further evaluate \mysys{} against a white-box adaptive attack that explicitly optimizes against the verification rule. For each evaluated step $t$, let $G_{i,t}^{\mathrm{GPU}}$ be the honest gradient. The attacker submits
\begin{equation}
\widetilde{G}_{i,t}
=
G_{i,t}^{\mathrm{GPU}}+\delta_t,
\qquad
\|\delta_t\|_2
=
\beta\|G_{i,t}^{\mathrm{GPU}}\|_2,
\end{equation}
where $\beta\in\{0.01,0.1,0.5,1,2,5,10\}$ controls the attack strength. The attacker has access to the model state, optimizer state, and verification rule, and searches for a perturbation that changes a previously correct prediction while remaining acceptable to the verifier. Specifically, it maximizes the post-update prediction loss together with a soft penalty for violating the corresponding verification constraint. The final candidate must still pass the actual verification rule; the optimization penalty itself does not determine acceptance.

We construct the attack separately for each verifier, so each method is evaluated against an adversary adapted to its own acceptance rule. We compare \mysys{} with RoFL-$L_2$ and RoFL-$L_\infty$~\citep{lycklama2023rofl}, EIFFeL-NormBall~\citep{chowdhury2022eiffel}, and RiseFL-Gaussian~\citep{zhu2024risefl}. These baselines are implemented as gradient-domain verification rules rather than full end-to-end reproductions of their cryptographic protocols. For \mysys{}, the attacker is optimized against the frozen deployed boundary and the same-step MPC reference $G_{i,t}^{\mathrm{MPC}}$. Detailed verifier definitions and attack parameters are provided in Appendix~\ref{app:adaptive-attack}.

We evaluate 200 instances per model for LeNet, BERT-Base, and Qwen3-0.6B, restricting evaluation to instances whose honest update produces the correct prediction. An attack succeeds only if the submitted gradient passes verification and changes that prediction from correct to incorrect. As summarized in Table~\ref{tab:attack-summary}(b), \mysys{} achieves 0\% ASR for every model and attack strength. In contrast, the baseline rules remain vulnerable to adaptive perturbations. Full results for all values of $\beta$ are reported in Appendix~\ref{app:adaptive-attack}. Together with the 0\% FRR above, these results indicate that \mysys{} separates the evaluated malicious deviations from honest computation rather than simply applying an overly restrictive acceptance rule.

We additionally evaluate tail-sparse PGD attacks on LeNet, restricting perturbations to the top 1\% or 10\% of coordinates ranked by absolute loss-gradient magnitude, with the same attack settings and strengths. The 1\% setting uses the full percentile grid, whereas the 10\% setting retains only checks at or below P90; both retain the infinity-norm guard. Without verification, ASR reaches 43.0\% and 56.5\%, respectively; with \mysys{}, it remains 0\% at every tested strength (Appendix~\ref{app:adaptive-attack}). These results support robustness in the evaluated settings, but do not establish security for arbitrary percentile grids or support sets. Adding percentile checks while retaining existing thresholds can further constrain accepted discrepancies, at additional verification cost.

\begin{table}[htbp]
\centering
\caption{Summary of attack evaluation. (a) General-attack results across all evaluated models and settings. (b) Maximum ASR (\%) over PGD attack strengths $\beta\in\{0.01,0.1,0.5,1,2,5,10\}$.}
\label{tab:attack-summary}
\begin{minipage}[t]{0.48\linewidth}
\centering

\caption*{(a) General attacks}
\setlength{\tabcolsep}{3pt}
\begin{tabular}{@{}lccc@{}}
\toprule
\textbf{Attack} & \textbf{Parameters} & \textbf{ASR} & \textbf{FRR} \\
\midrule
Gradient Reuse
& $k\in\{2,5,10\}$
& 0\% & 0\% \\
Reverse
& $\gamma\in\{0.5,1,2\}$
& 0\% & 0\% \\
Label Flip
& full replacement
& 0\% & 0\% \\
Amplify
& $\gamma\in\{5,10\}$
& 0\% & 0\% \\
\bottomrule
\end{tabular}
\end{minipage}
\hfill
\begin{minipage}[t]{0.48\linewidth}
\centering

\caption*{(b) Adaptive PGD}
\begin{tabular}{lccc}
\toprule
\textbf{Verifier} & \textbf{LeNet} & \textbf{BERT} & \textbf{Qwen3} \\
\midrule
No verifier
& 100.0 & 90.0 & 35.0 \\
RoFL-$L_2$
& 90.5 & 33.5 & 7.0 \\
RoFL-$L_\infty$
& 99.5 & 37.5 & 32.0 \\
EIFFeL
& 89.5 & 32.0 & 7.0 \\
RiseFL
& 97.99 & 34.5 & 12.42 \\
\textbf{\mysys{}}
& \textbf{0} & \textbf{0} & \textbf{0} \\
\bottomrule
\end{tabular}
\end{minipage}

\end{table}

\subsection{Stability and generalization of the empirical boundary}
\label{sec:boundary-generalization}
Finally, we study how many calibration steps are needed to obtain a stable boundary and whether the resulting boundary generalizes beyond the calibration setting. For BERT-Base, we calibrate on SST-2 with sequence length 32 and batch size 1 using the first $n\in\{20,50,100,200,300,500\}$ steps. Each resulting boundary is then frozen and evaluated without recalibration across held-out settings with different input sequence lengths, datasets, and GPUs.

\begin{figure}[htbp]
\centering

\begin{minipage}[t]{0.63\linewidth}
\vspace{0pt}
\centering

\setlength{\tabcolsep}{4pt}
\captionof{table}{BERT-Base FRR versus calibration steps.}
\label{tab:boundary-calibration}
\vspace{2pt}
\begin{tabular}{rccccc}
\toprule
\textbf{Steps} &
\textbf{SST-2/32} &
\textbf{SST-2/64} &
\textbf{CoLA/32} &
\textbf{A4500} &
\textbf{Cost} \\
\midrule
20  & 2.33\% & 32.40\% & 0.10\% & 2.00\% & 0.2$\times$ \\
50  & 0\%    & 4.10\%  & 0\%    & 0\%    & 0.5$\times$ \\
\textbf{100} &
\textbf{0\%} &
\textbf{0\%} &
\textbf{0\%} &
\textbf{0\%} &
1$\times$ \\
200 & 0\% & 0\% & 0\% & 0\% & 2$\times$ \\
300 & 0\% & 0\% & 0\% & 0\% & 3$\times$ \\
500 & 0\% & 0\% & 0\% & 0\% & 5$\times$ \\
\bottomrule
\end{tabular}
\end{minipage}
\hfill
\begin{minipage}[t]{0.33\linewidth}
\vspace{0pt}
\centering
\includegraphics[width=\linewidth]{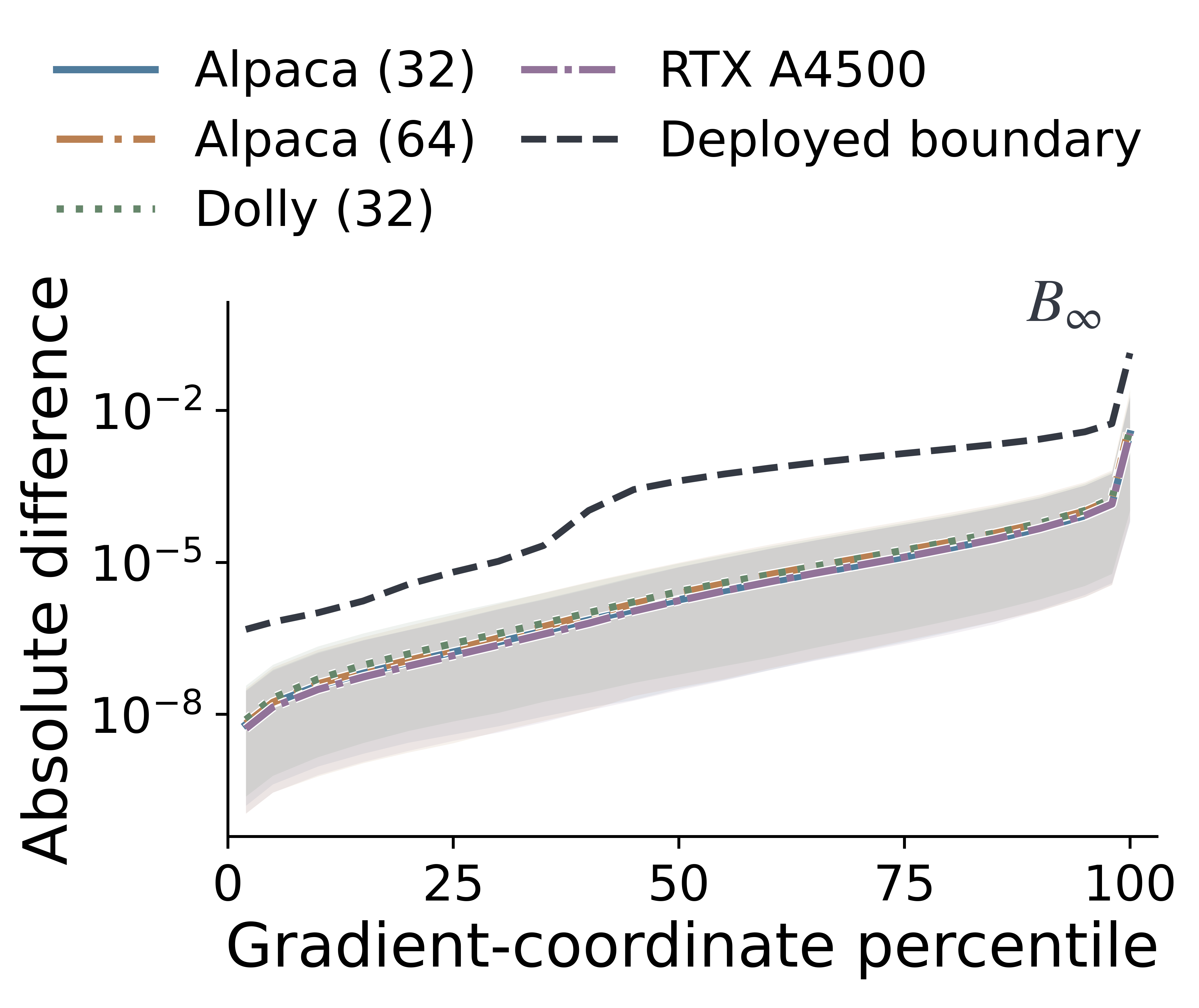}
\captionof{figure}{Generalization of the Qwen3 boundary.}
\label{fig:boundary-generalization}
\end{minipage}

\end{figure}

As shown in Table~\ref{tab:boundary-calibration}, small calibration sets can yield overly tight boundaries. With 20 steps, the FRR reaches 32.4\% on SST-2/64 and 2.0\% on the RTX A4500 setting; with 50 steps, SST-2/64 still has a 4.1\% FRR. In contrast, 100 calibration steps achieve 0\% FRR across all four evaluated settings, and increasing the calibration size to 200--500 steps provides no further improvement while increasing calibration cost. We therefore adopt 100 steps as the default calibration budget.

We observe the same trend for Qwen3 and LeNet: 100 calibration steps yield 0\% FRR across all evaluated generalization settings. Figure~\ref{fig:boundary-generalization} shows the generalization of Qwen3 absolute boundary and tail guard calibrated from 100 steps across different settings, with additional results in Appendix~\ref{app:boundary-generalization}. These results support the stability and generalization of the calibrated boundary.

\subsection{System Overhead and Cost Amortization}
\label{sec:system-overhead}

This section evaluates the computational cost of \mysys{} and the trade-off between audit rate and economic deterrence. Our goal is to quantify how optimistic auditing amortizes expensive private verification, rather than to provide a strict system-level comparison across different cryptographic backends.

We construct a five-client FedSGD workload using LeNet with 0.0617M parameters, following the configuration evaluated in ZKSL~\citep{zheng2026zksl}. Each client performs one local gradient step per round with batch size 1, and the server equally aggregates the five gradients. We consider 1,000 rounds, corresponding to 5,000 client gradient computations and 1,000 aggregations.

We compare ZKSL-G, ZKSL-LP-G, full MPC-based FL, and \mysys{} with audit rates $p\in\{0.10,0.05,0.01\}$. ZKSL costs are extrapolated from the original work, while MPC costs are estimated from our SPU implementation. The reported time includes verification and aggregation but excludes initialization, communication, commitment, and one-time boundary calibration costs.

\begin{table}[t]
\centering

\setlength{\tabcolsep}{4pt}
\caption{Verification cost and normalized deposit. K denotes $10^3$. Deposit ratios are normalized to \mysys{} with $p=0.10$ and use $p_{\mathrm{FR}}=0.01$.}
\label{tab:system-cost}
\begin{tabular}{lrrrr}
\toprule
Method &
Time (s) &
Time ratio &
Deposit &
Deposit ratio \\
\midrule
ZKSL-G             & 16.09K & 8.38 & 0.00 & 0.00 \\
ZKSL-LP-G          & 12.19K & 6.35 & 0.00 & 0.00 \\
Full MPC-based FL  & 1.92K  & 1.00 & 0.00 & 0.00 \\
\mysys{}, $p=0.10$ & 192.35 & 0.10 & 1.11 & 1.00 \\
\mysys{}, $p=0.05$ & 96.32  & 0.05 & 1.12 & 1.01 \\
\mysys{}, $p=0.01$ & 19.49  & 0.01 & 1.76 & 1.59 \\
\bottomrule
\end{tabular}
\end{table}

Table~\ref{tab:system-cost} shows that the cost of \mysys{} decreases nearly linearly with the audit rate. At $p=0.01$, \mysys{} is approximately $98.6\times$ faster than full MPC-based FL and $625.5\times$ faster than ZKSL-LP-G under this workload. These results should be interpreted as a cost analysis rather than a strict end-to-end speedup because the measurements use different hardware environments. They illustrate that \mysys{} pays the cost of expensive private replay only for a small fraction of committed computations.

\paragraph{Audit rate and deposit trade-off.}
A lower audit rate reduces verification cost but also decreases the probability of detecting deviations. Consider a client that skips 100 of its 1,000 rounds, with the saved computation normalized to one unit. If each round is independently audited with probability $p$, the probability of detecting at least one skipped computation is
$\Pr_{\mathrm{det}}(p)=1-(1-p)^{100}$.

Because the verification boundary is empirically calibrated, we conservatively allow an overall false-rejection probability $p_{\mathrm{FR}}=0.01$ for an honest client. Let $\mathsf{Stake}(p)$ denote the refundable deposit. Relative to honest execution, the expected additional utility of deviation is
$\Delta U(p)=1-(\Pr_{\mathrm{det}}(p)-p_{\mathrm{FR}})\mathsf{Stake}(p)$.
We use a 10\% deterrence margin and set
$\mathsf{Stake}(p)=1.1/(\Pr_{\mathrm{det}}(p)-p_{\mathrm{FR}})$,
which gives $\Delta U(p)=-0.1$. As shown in Table~\ref{tab:system-cost}, reducing $p$ from $0.10$ to $0.05$ has little effect on the required deposit, while $p=0.01$ increases it to $1.59\times$ the deposit required at $p=0.10$.

\paragraph{Deposit considerations.}
This analysis illustrates the trade-off between computational auditing and economic deterrence rather than prescribing a universal deposit value. Malicious behavior may provide benefits beyond saved computation and can be difficult to quantify. Deployed blockchain systems therefore often require nontrivial collateral and penalize provable misbehavior~\citep{ethereumpos,arbitrumnitro}. For example, Ethereum validators are subject to stake slashing, while Arbitrum uses stake-backed claims and challenge penalties. A deployment of \mysys{} can therefore choose the deposit conservatively according to the expected attack benefit and measured reliability of the deployed boundary.
\section{Conclusion}
We presented \mysys{}, an optimistic verification framework for privacy-preserving federated learning. \mysys{} combines private MPC replay with an empirical gradient-discrepancy boundary to verify sampled client computations without requiring bitwise agreement across execution environments. Our experiments show that the boundary remains stable across diverse training settings and detects the evaluated direct and adaptive attacks while maintaining a low FRR. By applying an optimistic verification mechanism, \mysys{} also substantially reduces verification cost. We believe this provides a practical direction for verifying federated learning while preserving data privacy.

\subsection*{AI use statement} We used generative AI tools to improve the clarity and readability of the manuscript, assist with literature searches, and support code development and debugging. The authors reviewed and revised the AI-assisted text, checked the cited sources for accuracy and relevance, and reviewed and tested the AI-assisted code. The authors take responsibility for the final manuscript, including all claims, references, code, and reported results.

\bibliography{iclr2027_conference}
\bibliographystyle{iclr2027_conference}

\appendix
\section{Additional Boundary Generalization Results}
\label{app:boundary-generalization}

\begin{figure}[htbp]
    \centering

    \begin{subfigure}[t]{0.32\linewidth}
        \centering
        \includegraphics[width=\linewidth]{figs/qwen_absolute_tail_guard.png}
        \caption{Qwen3: absolute}
        \label{fig:qwen-absolute}
    \end{subfigure}
    \hfill
    \begin{subfigure}[t]{0.32\linewidth}
        \centering
        \includegraphics[width=\linewidth]{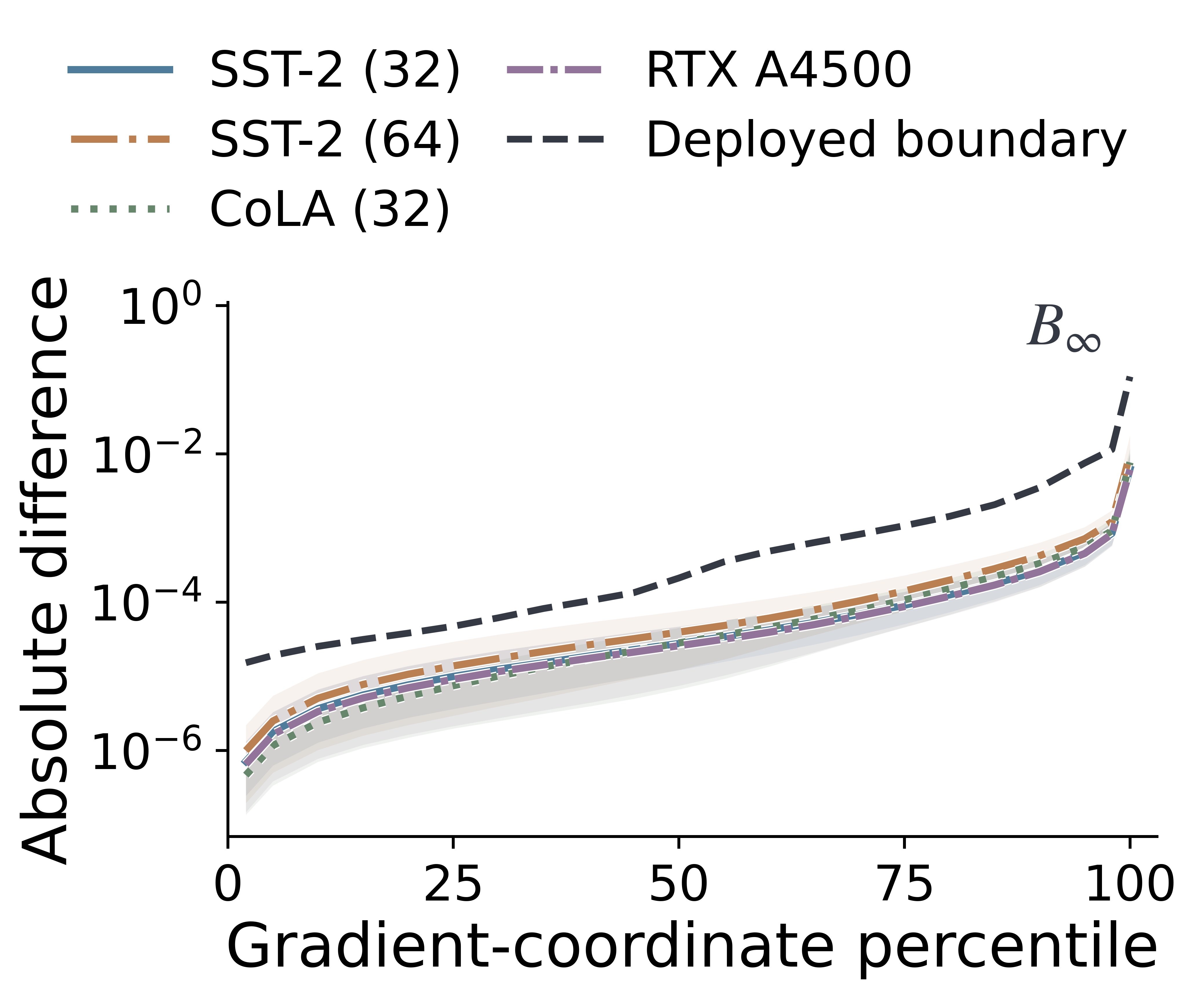}
        \caption{BERT: absolute}
        \label{fig:bert-absolute}
    \end{subfigure}
    \hfill
    \begin{subfigure}[t]{0.32\linewidth}
        \centering
        \includegraphics[width=\linewidth]{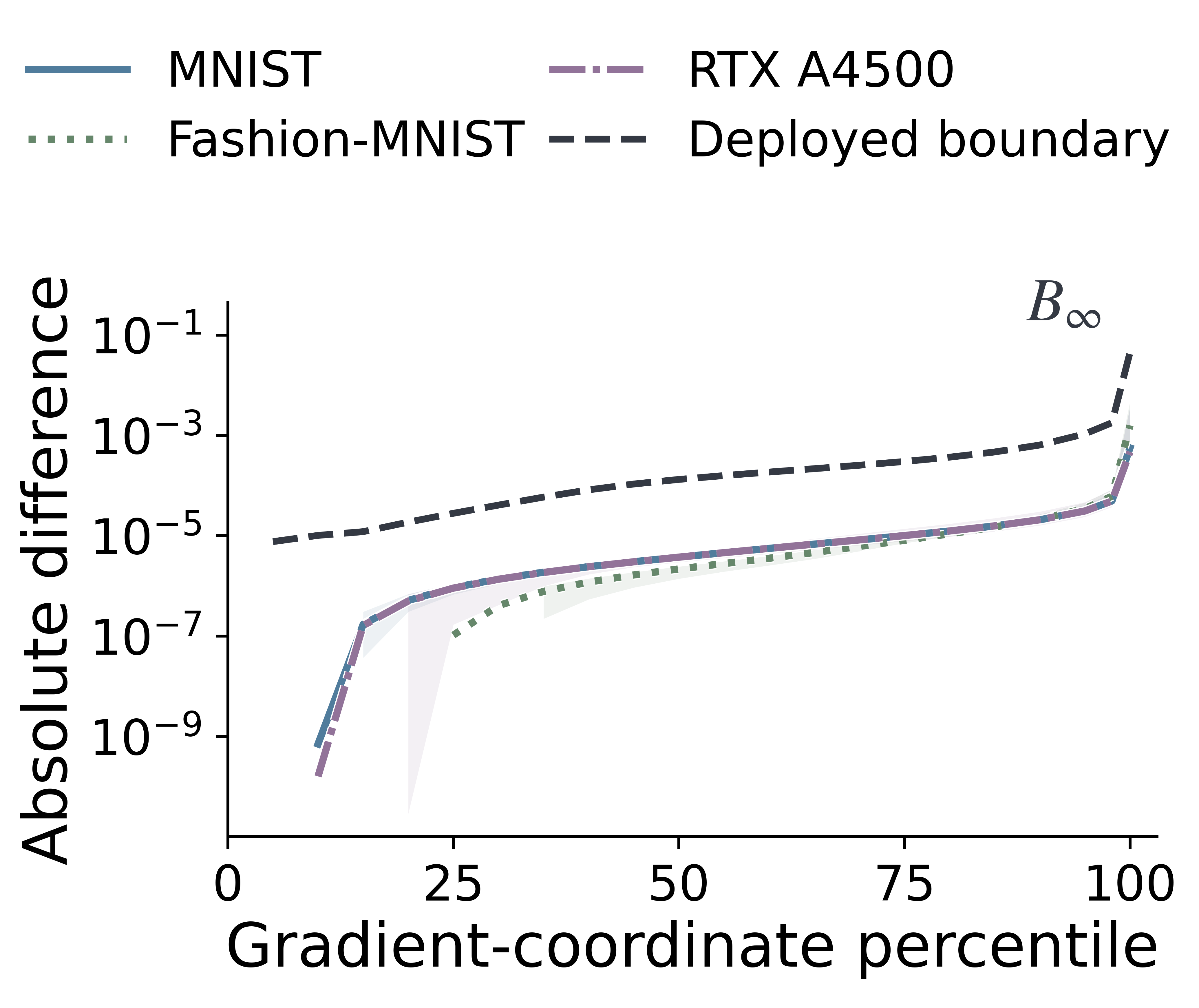}
        \caption{LeNet: absolute}
        \label{fig:lenet-absolute}
    \end{subfigure}

    \par\medskip

    \begin{subfigure}[t]{0.32\linewidth}
        \centering
        \includegraphics[width=\linewidth]{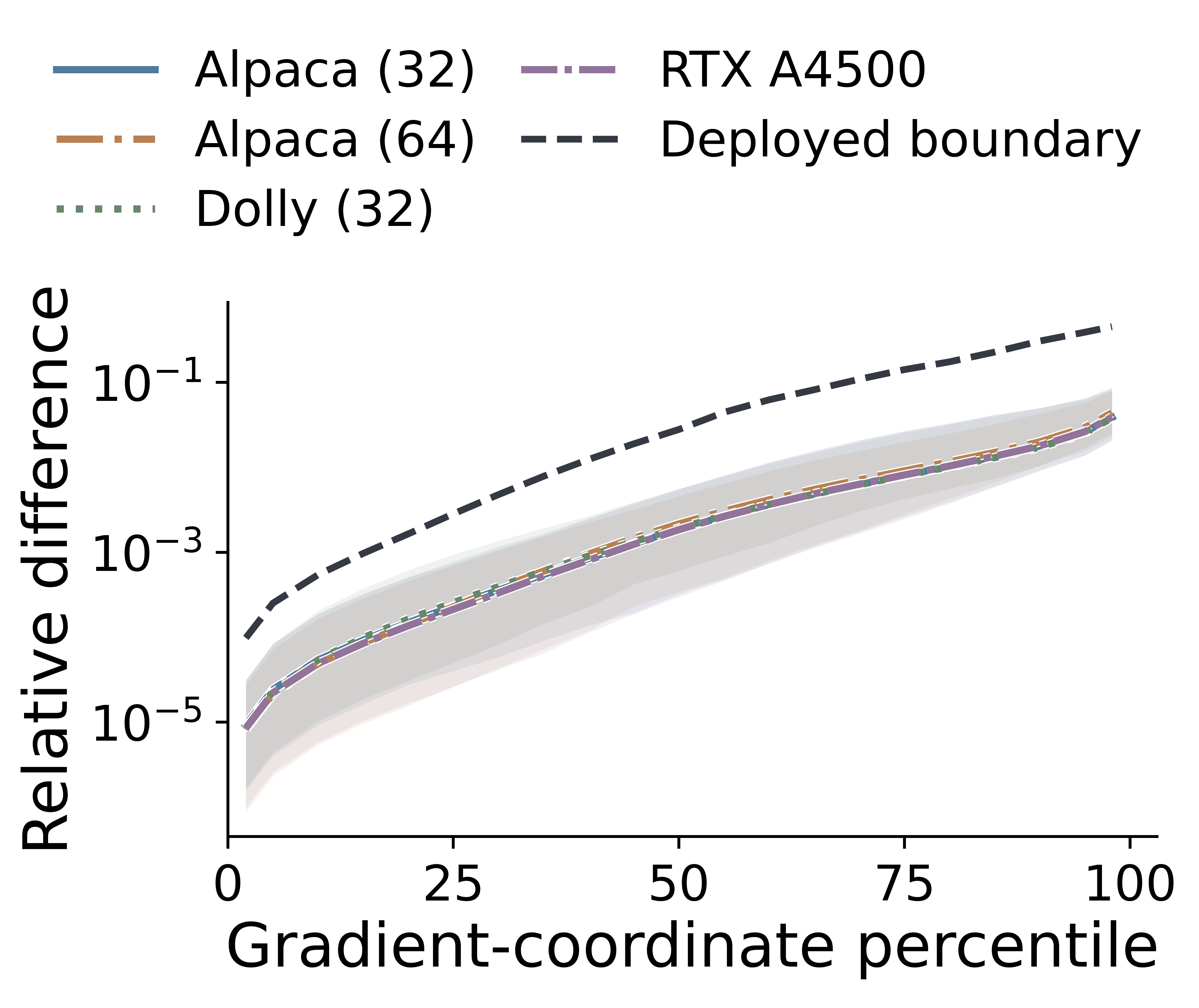}
        \caption{Qwen3: relative}
        \label{fig:qwen-relative}
    \end{subfigure}
    \hfill
    \begin{subfigure}[t]{0.32\linewidth}
        \centering
        \includegraphics[width=\linewidth]{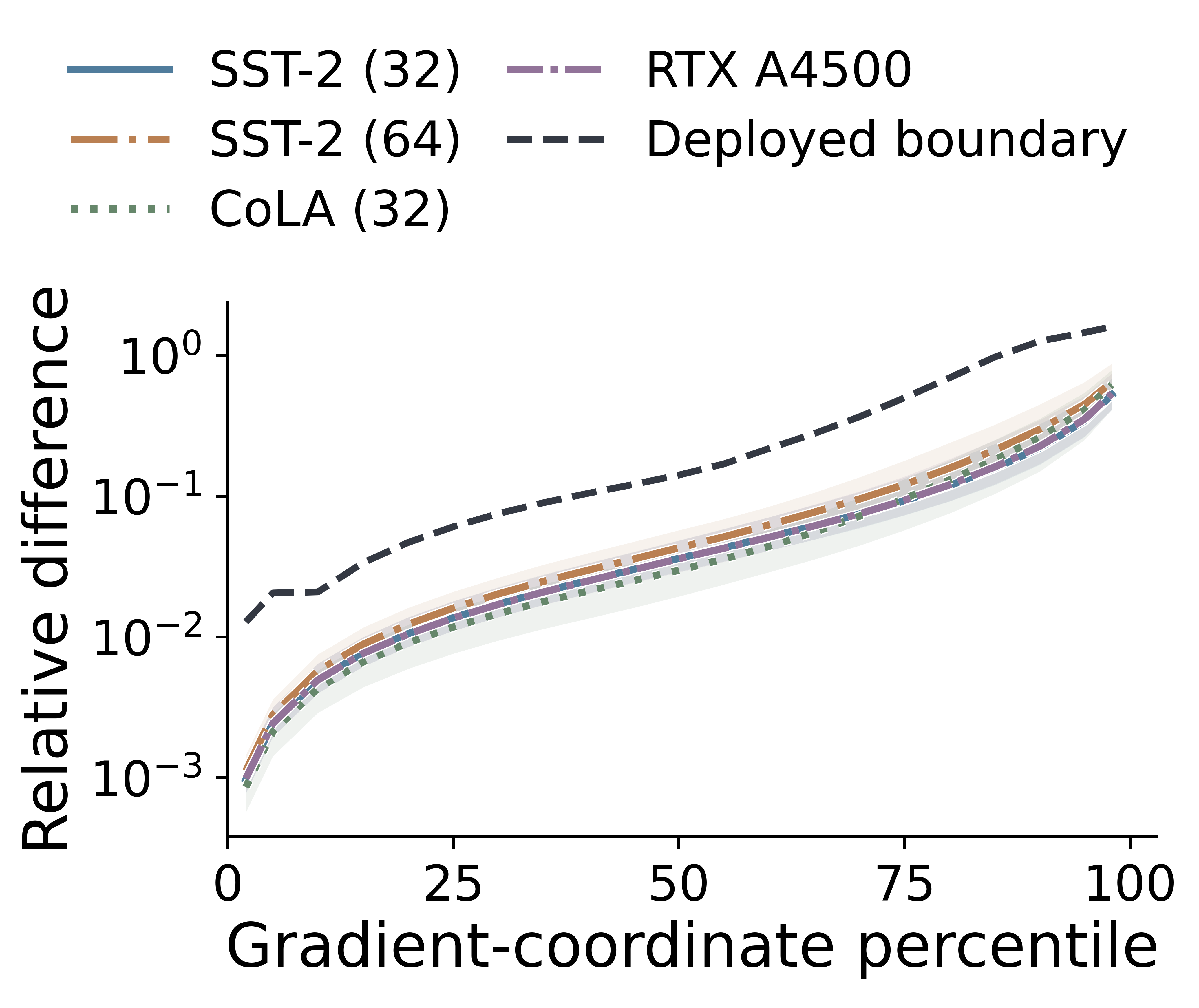}
        \caption{BERT: relative}
        \label{fig:bert-relative}
    \end{subfigure}
    \hfill
    \begin{subfigure}[t]{0.32\linewidth}
        \centering
        \includegraphics[width=\linewidth]{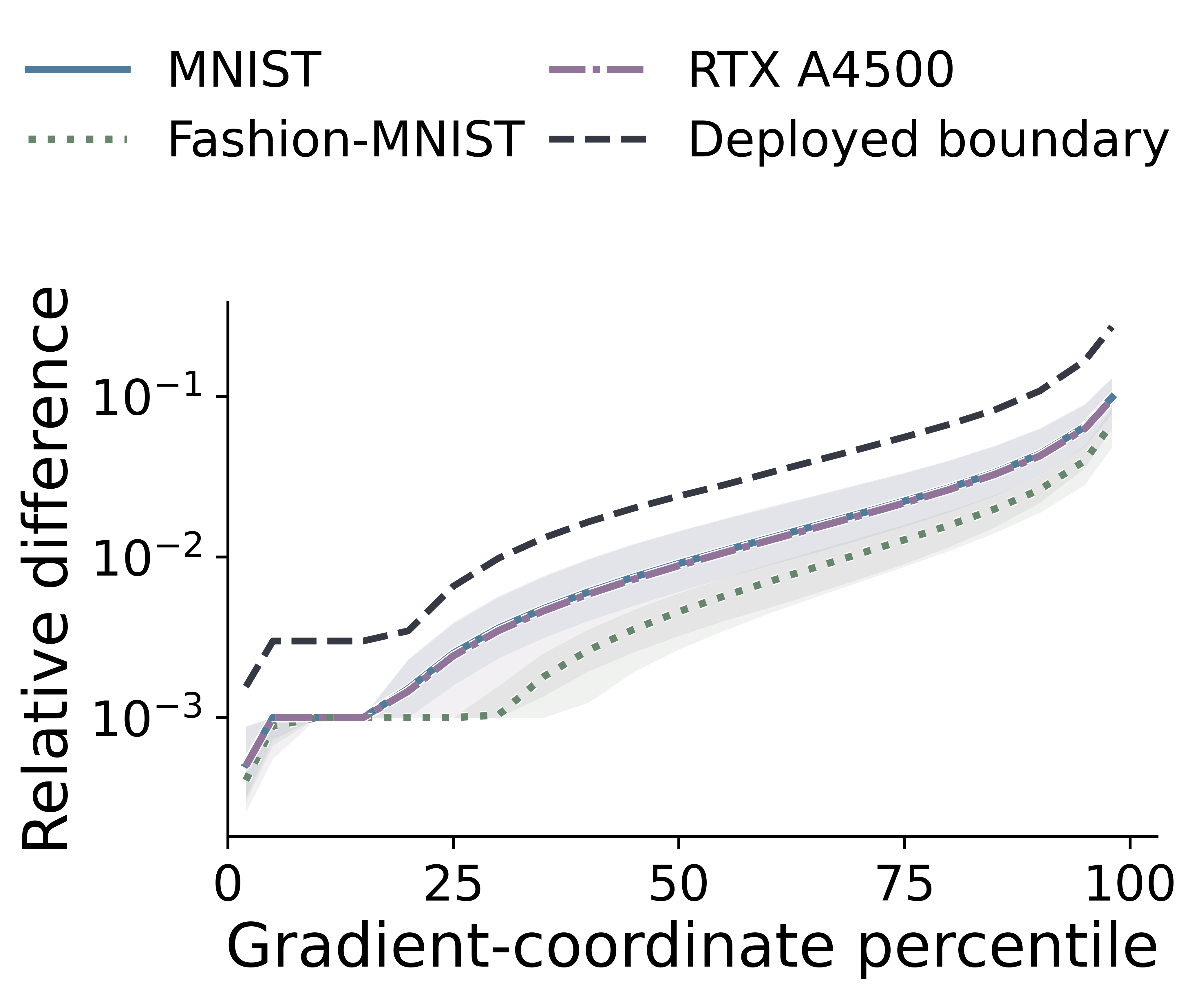}
        \caption{LeNet: relative}
        \label{fig:lenet-relative}
    \end{subfigure}

    \caption{Generalization of the deployed boundaries calibrated from 100 steps for Qwen3, BERT, and LeNet. The top and bottom rows show absolute and relative gradient discrepancies, respectively. Solid lines and shaded regions show the median and percentile range across evaluation steps, respectively; dashed lines indicate the deployed boundaries.}
    \label{fig:boundary-generalization-all}
\end{figure}

Figure~\ref{fig:boundary-generalization-all} complements Figure~\ref{fig:boundary-generalization} by providing a visual comparison of the boundary and gradient differences for all three models: Qwen3, LeNet, and BERT-Base. The deployed boundary is calibrated using the first 100 training steps and is then frozen for evaluation. We use a safety factor of $\alpha=4$ for Qwen3 and $\alpha=3$ for LeNet. Across all evaluated generalization settings, the resulting boundaries achieve 0\% FRR, consistent with the trend observed for BERT-Base.

\section{Adaptive PGD Attack Details}
\label{app:adaptive-attack}

\paragraph{Attack construction.}
We evaluate a white-box adaptive PGD attack against each verification rule. For an evaluated client-step pair $(i,t)$, let $G_{i,t}^{\mathrm{GPU}}$ denote the honest gradient. The attacker submits
\begin{equation}
\widetilde{G}_{i,t}
=
G_{i,t}^{\mathrm{GPU}}+\delta_t,
\qquad
\|\delta_t\|_2
=
\beta
\|G_{i,t}^{\mathrm{GPU}}\|_2,
\end{equation}
where
\[
\beta\in\{0.01,0.1,0.5,1,2,5,10\}.
\]

Let $U_t(G)$ denote the model parameters obtained by applying gradient $G$ from the saved model parameters and optimizer state at step $t$. The attacker searches for a perturbation that maximizes the post-update prediction loss while remaining compatible with the target verifier:
\begin{equation}
\max_{\|\delta\|_2=\beta\|G_{i,t}^{\mathrm{GPU}}\|_2}
\ell\!\left(
f_{U_t(G_{i,t}^{\mathrm{GPU}}+\delta)}(x_t),
y_t
\right)
-
\lambda
\mathcal{P}_V
\left(
G_{i,t}^{\mathrm{GPU}}+\delta
\right),
\end{equation}
where $\mathcal{P}_V$ is a differentiable penalty associated with verifier $V$ and $\lambda=100$.

We optimize each verifier separately rather than generating a single unconstrained attack and evaluating it against all methods. Optimization uses normalized gradient ascent for 20 iterations. After each iteration, the perturbation is projected back to the fixed-radius sphere. The step size is set to one fifth of the perturbation radius. Each attack uses three initializations: one aligned with the prediction-loss gradient and two random directions. We use random seed 42. Importantly, the soft penalty is used only to guide optimization; the final candidate must still pass the actual verification rule.

We evaluate 200 instances for each model. All selected instances are required to produce a correct prediction after the honest update. For LeNet and BERT, an attack succeeds when the post-update class prediction changes from correct to incorrect. For Qwen3, we evaluate the final supervised token position and require the originally correct token prediction to become incorrect. Thus, the attack objective is an untargeted prediction flip.

\paragraph{Verification baselines.}
We compare \mysys{} against gradient-domain adaptations of RoFL-$L_2$, RoFL-$L_\infty$, EIFFeL-NormBall, and RiseFL-Gaussian, together with a no-verification baseline. These experiments compare the corresponding gradient acceptance rules rather than reproducing the complete end-to-end cryptographic protocols.

Using the first 100 honest training steps, we define
\begin{equation}
B_2
=
\max_{t<100}
\left\|
G_{i,t}^{\mathrm{GPU}}
\right\|_2,
\qquad
B_{\infty}^{\mathrm{RoFL}}
=
\max_{t<100}
\left\|
G_{i,t}^{\mathrm{GPU}}
\right\|_\infty,
\end{equation}
and
\begin{equation}
R
=
\max_{t<100}
\left\|
G_{i,t}^{\mathrm{GPU}}
-
V_t
\right\|_2,
\end{equation}
where $V_t$ is a reference gradient computed from a fixed public reference dataset at the saved model state.

\begin{table}[htbp]
\centering
\caption{Verification rules used in the adaptive PGD evaluation.}
\label{tab:adaptive-verifiers}
\begin{tabular}{ll}
\toprule
\textbf{Verifier} & \textbf{Acceptance rule} \\
\midrule
No verifier
& Always accept \\
RoFL-$L_2$
& $\|\widetilde{G}_{i,t}\|_2\leq B_2$ \\
RoFL-$L_\infty$
& $\|\widetilde{G}_{i,t}\|_\infty\leq B_{\infty}^{\mathrm{RoFL}}$ \\
EIFFeL-NormBall
& $\|\widetilde{G}_{i,t}-V_t\|_2\leq R$ \\
RiseFL-Gaussian
& Gaussian projection norm test \\
\mysys{}
& $P_{i,t}^{\mathrm{abs}}(p)\leq B_p^{\mathrm{abs}}$
and
$P_{i,t}^{\mathrm{rel}}(p)\leq B_p^{\mathrm{rel}}$,
$\forall p\in\mathcal{P}$ and
$\Delta_{i,t}^{\infty}\le B_\infty$ \\
\bottomrule
\end{tabular}
\end{table}

The corresponding acceptance rules are summarized in Table~\ref{tab:adaptive-verifiers}. All constraints are applied to the submitted gradient $\widetilde{G}_{i,t}$ rather than to the perturbation $\delta_t$.

For the deterministic norm-based baselines, the adaptive penalty is defined as
\begin{equation}
\mathcal{P}_V(u)
=
\left[
\max
\left(
\frac{c_V(u)}{b_V}-1,
0
\right)
\right]^2,
\end{equation}
where $c_V(u)$ denotes the corresponding norm statistic and $b_V$ its threshold.

For RiseFL-Gaussian, we use $K=1000$ Gaussian projections and $\epsilon=2^{-128}$. For a fixed submitted gradient $u$, the quantile $q$, the analytical acceptance probability $p_{\mathrm{acc}}(u)$ of $u$, and the penalty $\mathcal{P}_V(u)$ used in the adaptive objective are defined as
\begin{equation}
\begin{aligned}
q &= F_{\chi^2_K}^{-1}(1-\epsilon),\\
p_{\mathrm{acc}}(u)
  &= F_{\chi^2_K}\left(\frac{qB_2^2}{\|u\|_2^2}\right),\\
\mathcal{P}_V(u) &= \left(1-p_{\mathrm{acc}}(u)\right)^2.
\end{aligned}
\end{equation}
After optimization, the selected candidate is evaluated using an independent randomized audit.

For \mysys{}, the submitted gradient is compared with the same-step MPC replay $G_{i,t}^{\mathrm{MPC}}$. The deployed empirical boundary is frozen before the attack and is not recalibrated using adversarial samples. We use safety factors $(\alpha_{\mathrm{abs}},\alpha_{\mathrm{rel}},\alpha_{\infty}) =(3,3,3)$ for LeNet, $(3,3,6)$ for BERT, and $(4,4,4)$ for Qwen3.

\paragraph{Attack success rate.}
An attack is considered successful only if it both passes verification and changes an originally correct prediction to an incorrect one. For $N=200$ evaluated instances, we report
\begin{equation}
\mathrm{ASR}
=
\frac{1}{N}
\sum_{n=1}^{N}
\mathbf{1}
\left[
\text{prediction flip}
\right]
\,
p_{\mathrm{acc}}
\left(
\widetilde{G}_{i,t}^{(n)}
\right).
\end{equation}
For deterministic verifiers,
$p_{\mathrm{acc}}\in\{0,1\}$. For RiseFL, we report the expected ASR weighted by its analytical acceptance probability. If optimization fails to find an acceptable malicious candidate and falls back to the honest gradient, the instance is counted as an attack failure and remains in the denominator.

\paragraph{Full results.}
Tables~\ref{tab:pgd-lenet}, \ref{tab:pgd-bert}, and \ref{tab:pgd-qwen} report the complete results for all tested attack strengths.

\begin{table}[H]
\centering
\caption{Adaptive PGD attack success rate (\%) on LeNet.}
\label{tab:pgd-lenet}
\begin{tabular}{rcccccc}
\toprule
$\beta$
& \textbf{None}
& \textbf{RoFL-$L_2$}
& \textbf{RoFL-$L_\infty$}
& \textbf{EIFFeL}
& \textbf{RiseFL}
& \textbf{\mysys{}} \\
\midrule
0.01 & 0.0   & 0.0  & 0.0  & 0.0  & 0.0   & 0.0 \\
0.1  & 4.5   & 4.5  & 4.5  & 4.5  & 4.5   & 0.0 \\
0.5  & 29.5  & 29.5 & 29.5 & 29.5 & 29.5  & 0.0 \\
1    & 57.0  & 57.0 & 57.0 & 57.0 & 57.0  & 0.0 \\
2    & 90.0  & 90.0 & 90.0 & 89.5 & 90.0  & 0.0 \\
5    & 99.0  & 90.5 & 99.0 & 68.0 & 97.99 & 0.0 \\
10   & 100.0 & 37.0 & 99.5 & 13.0 & 71.48 & 0.0 \\
\bottomrule
\end{tabular}
\end{table}

\begin{table}[H]
\centering
\caption{Adaptive PGD attack success rate (\%) on BERT-Base.}
\label{tab:pgd-bert}
\begin{tabular}{rcccccc}
\toprule
$\beta$
& \textbf{None}
& \textbf{RoFL-$L_2$}
& \textbf{RoFL-$L_\infty$}
& \textbf{EIFFeL}
& \textbf{RiseFL}
& \textbf{\mysys{}} \\
\midrule
0.01 & 0.0  & 0.0  & 0.0  & 0.0  & 0.0  & 0.0 \\
0.1  & 4.5  & 4.5  & 4.5  & 4.5  & 4.5  & 0.0 \\
0.5  & 20.0 & 20.0 & 20.0 & 20.0 & 20.0 & 0.0 \\
1    & 35.5 & 33.5 & 35.0 & 32.0 & 34.5 & 0.0 \\
2    & 49.5 & 33.5 & 37.5 & 27.0 & 21.0 & 0.0 \\
5    & 72.0 & 0.0  & 14.0 & 0.0  & 0.0  & 0.0 \\
10   & 90.0 & 0.0  & 0.5  & 0.0  & 0.0  & 0.0 \\
\bottomrule
\end{tabular}
\end{table}

\begin{table}[H]
\centering
\caption{Adaptive PGD attack success rate (\%) on Qwen3-0.6B.}
\label{tab:pgd-qwen}
\begin{tabular}{rcccccc}
\toprule
$\beta$
& \textbf{None}
& \textbf{RoFL-$L_2$}
& \textbf{RoFL-$L_\infty$}
& \textbf{EIFFeL}
& \textbf{RiseFL}
& \textbf{\mysys{}} \\
\midrule
0.01 & 0.0  & 0.0 & 0.0  & 0.0 & 0.0  & 0.0 \\
0.1  & 0.0  & 0.0 & 0.0  & 0.0 & 0.0  & 0.0 \\
0.5  & 0.0  & 0.0 & 0.0  & 0.0 & 0.0  & 0.0 \\
1    & 0.0  & 0.0 & 0.0  & 0.0 & 0.0  & 0.0 \\
2    & 7.0  & 7.0 & 7.0  & 7.0 & 7.0  & 0.0 \\
5    & 20.0 & 5.5 & 20.0 & 5.5 & 12.42 & 0.0 \\
10   & 35.0 & 1.0 & 32.0 & 1.0 & 2.99  & 0.0 \\
\bottomrule
\end{tabular}
\end{table}

Across all three models, \mysys{} maintains 0\% ASR for every tested attack strength. The norm-based baselines reject some large perturbations, but remain vulnerable to adaptive attacks that satisfy their respective acceptance constraints, particularly at moderate perturbation strengths.

\paragraph{Tail-sparse PGD attacks.}
Table~\ref{tab:pgd-tail-lenet} reports the LeNet results for perturbations restricted to the top 1\% or 10\% of coordinates ranked by absolute loss-gradient magnitude, using the attack settings and strengths described above. This experiment uses independently specified verification grids. For the 1\% support setting, we use $\mathcal{P}_{1\%}=\{0.02,0.05,0.10,0.15,\ldots,0.90,0.95,0.98,0.99\}$. For the 10\% support setting, we use $\mathcal{P}_{10\%}=\{p\in\mathcal{P}_{1\%}:p\leq0.90\}$. Both settings retain the infinity-norm guard $B_\infty$. 

\begin{table}[htbp]
\centering
\caption{Tail-sparse PGD attack success rate (\%) on LeNet.}
\label{tab:pgd-tail-lenet}
\begin{tabular}{rcccc}
\toprule
& \multicolumn{2}{c}{\textbf{1\% support}} & \multicolumn{2}{c}{\textbf{10\% support}} \\
\cmidrule(lr){2-3}\cmidrule(lr){4-5}
$\beta$ & No verifier & \mysys{} & No verifier & \mysys{} \\
\midrule
0.01 & 0.0  & 0.0 & 0.0  & 0.0 \\
0.1  & 0.0  & 0.0 & 0.0  & 0.0 \\
0.5  & 0.0  & 0.0 & 0.0  & 0.0 \\
1    & 1.5  & 0.0 & 4.5  & 0.0 \\
2    & 6.5  & 0.0 & 12.0 & 0.0 \\
5    & 24.5 & 0.0 & 28.5 & 0.0 \\
10   & 43.0 & 0.0 & 56.5 & 0.0 \\
\bottomrule
\end{tabular}
\end{table}

\end{document}